\documentclass[a4paper,11pt]{article}
\usepackage{graphicx}
\usepackage{float}
\date{}
\usepackage[utf8x]{inputenc}

\title{Theoretical analysis of neutron scattering results for quasi-two dimensional ferromagnets}
\author{Subhajit Sarkar 
\footnote{email: subhajit@bose.res.in}, Samir K. Paul \footnote{email: smr@bose.res.in}  and Ranjan Chaudhury \footnote{email: ranjan@bose.res.in} \\
S N Bose National Centre for Basic Sciences, \\ Block- JD, Sector- III, Salt Lake, Kolkata- 700098, India.
}
\begin{document}
\maketitle
\begin{abstract}
A theoretical study has been carried out to analyze the available results from the inelastic neutron scattering experiment performed on a quasi-two dimensional spin-$\frac{1}{2}$ ferromagnetic material $K_2CuF_4$. Our formalism is based on a conventional semi-classical like treatment involving a model of an ideal gas of vortices/anti-vortices corresponding to an anisotropic XY Heisenberg ferromagnet on a square lattice. The results for dynamical structure functions for our model corresponding to spin-$\frac{1}{2}$, show occurrence of negative values in a large range of energy transfer even encompassing the experimental range, when convoluted with a realistic spectral window function. This result indicates failure of the conventional theoretical framework to be applicable to the experimental situation corresponding to low spin systems. A full quantum formalism seems essential for treating such systems.
\end{abstract}
\begin{center}
\\ 	75.30.Kz--Kosterlitz-Thouless transition in magnetic systems
\end{center}
%
%
\section{Introduction}
\paragraph*{}
Low dimensional and in particular two dimensional magnetism has attracted a great deal of interest in the past three decades \cite{1,2,3}. In particular, in one dimension the existence of both solitonic and spin wave excitations were thoroughly studied through inelastic neutron scattering experiments as well as theoretical analysis for $CsNiF_3$ \cite{4}. Similar studies were carried out searching for topological excitations in various quasi-one dimensional systems which are almost ideal realization of nearest neighbour Heisenberg antiferromagnetic chain \cite{5}.
\paragraph*{}
In many of the above systems the experiments showed central peak (peak corresponding to $\omega = 0$) in the dynamical structure function when plotted in constant $``q"$ scan. This motivated the experimentalists further to investigate two dimensional and quasi-two dimensional magnetic materials. With the availability of improved quasi two-dimensional ferromagnetic and anti-ferromagnetic materials investigations along this line has become todays one of the primary interests both theoretical and experimental. These include layered systems such as $K_2 Cu F_4$, $Rb_2 CrCl_4$, magnetically intercalated graphites such as $CoCl_2$, layered ruthenates, layered manganites and high $T_c$ cuprates \cite{1,2,3,4,6,7,8,9,10,11,12}. Moreover large amount of information on the spin dynamics, extracted from inelastic neutron scattering are available. Advances in numerical and computational techniques have also contributed to the understanding of both spin wave and topological excitations \cite{13,14,15,16,17}. 
\paragraph*{}
On two dimensional magnetic systems the concept of topological order was introduced by Kosterlitz and Thouless and independently by Berezinskii \cite{18,19}. Their ideas back-ed by analytical and numerical calculations led to the proposal for the existence of topological vortices and anti-vortices in a typical ferromagnetic XY model on a two dimensional lattice. According to these ideas vortices and anti-vortices are frozen as bound pairs below certain transition temperature called $T_{KT}$ or $T_{BKT}$ and above this temperature they become mobile and nearly free \cite{18,19}.
\paragraph*{}
In this work we initiate a theoretical investigation regarding the applicability of a semi-classical like treatment of the dynamics of topological excitations to the inelastic neutron scattering results for real systems \cite{6,13,14,20}. In recent years inelastic neutron scattering experiments are being done mostly on layered ruthenates like $Ca_{2-x}Sr_{x}RuO_{4}$, layered anti-ferromagnet like $CuGeO_{3}$, layered manganites  and some layered cuprates \cite{1,2,3,10,11,21}. The layered anti-ferromagnet, $CuGeO_{3}$ being spin-Peierls compound, is proposed to exhibit a Berezinskii-Kosterlitz-Thouless transition in the vicinity of spin-Peierls transition temperature. The two dimensional spin half XY model was investigated and the validity of Berezinskii-Kosterlitz-Thouless transition was confirmed \cite{22}. The existence of Berezinskii- Kosterlitz- Thouless transition was proposed long ago in $K_2CuF_4$(S= $\frac{1}{2}$ layered ferromagnet) \cite{6}. According to our knowledge layered ferromagnets with spin-$\frac{1}{2}$ are the least studied systems, both from theoretical and experimental point of view, till date.
\paragraph*{}
An extensive experimental study of spin-dynamics in a layered ferromagnet has been carried out by Hirakawa et. al. \cite{6} using neutron scattering probe on $K_2 CuF_4$.  Their results exhibit a central peak (at $\omega = 0$) in the plot of ``neutron count vs. frequency'' at a fixed value of the wave-vector $q$. Subsequent developments of approximate analytical theories and Monte Carlo Molecular Dynamics(MCMD) analysis have suggested that the existence of central peaks is partly due to scattering of neutrons from moving vortices and anti-vortices \cite{14,15}.
\paragraph*{}
Here we aim to examine how far the picture of ideal gas of vortices and anti-vortices could be extended to the quantum spin models. For this purpose, we choose $K_2 CuF_4$ as the reference system. This is a spin-$\frac{1}{2}$ quasi-two-dimensional ferromagnetic material, on which extensive neutron scattering studies have been done. 
\paragraph*{}
The plan of the paper is as follows. In Sec. 2, we briefly describe the classical theory of mobile vortices and anti-vortices. In the same Section we explain our mathematical formulations in detail. In Sec. 3 we discuss our calculations and results. In Sec. 4 we present the conclusions and the future plan.
 \section{Mathematical Formulation}
\paragraph*{}
The dynamics of mobile vortices in a ferromagnetic system has already been treated both analytically and numerically by Huber \cite{13} and Mertens et. al. \cite{14}. In this work we apply their classical formalism to study the phase transition in $K_2 CuF_4$. We calculate the spin-spin correlations taking the experimental situations into account. For our purpose we present a brief description of the analytical treatment developed in Ref. \cite{14}. The starting Hamiltonian is,
\begin{equation}
 \mathcal{H}= -J\sum_{\langle ij \rangle}(S_{i}^{x}S_{j}^{x}+S_{i}^{y}S_{j}^{y}+\lambda S_{i}^{z}S_{j}^{z}),
\end{equation}
where $i,j$ label the nearest neighbour sites on a two dimensional square lattice, J is the coupling constant and the classical spin vector is $\mathbf{S}_{i}\equiv (S_{i}^{x}, S_{i}^{y}, S_{i}^{z})$. This is an anisotropic Heisenberg    Hamiltonian which, for $J > 0$, represents a ferromagnetic system. The quantity $\lambda$ is the anisotropy parameter whose XY and isotropic Heisenberg limit correspond to $\lambda=0$ and $1$ respectively. 
The general time dependent spin configuration in spherical polar coordinate system is given by,
\begin{eqnarray}
 S_{x}&=& S\,cos \phi(\mathbf{r},t)sin \theta(\mathbf{r},t),\nonumber \\
 S_{y}&=& S\,sin \phi(\mathbf{r},t)sin \theta(\mathbf{r},t),\nonumber \\
 S_{z}&=& S\,cos \theta(\mathbf{r},t),
\end{eqnarray}
with $\mathbf{r}= (x,y)$. Following the formulation of Hikami and Tsuneto, the solutions are given by \cite{20}, $\phi= \pm \arctan (\frac{y}{x})$ and 
\begin{eqnarray}
\theta &=& \frac{\pi}{2}(1 \pm e^{-r/r_{v}})  \,\,\,\,\,\,\, for \,\,\,\,r\gg r_{v},\nonumber\\
       &=& 0 \,\,\,\,or \,\,\,\, \pi \,\,\,\,\,\,\, r\rightarrow 0,
\end{eqnarray}
for single vortex centred at $\mathbf{r}= (0,0)$, where (3) describes the asymptotic behaviour of $\theta$. Here vortex core radius is given by \cite{15}, $r_{v}\,=\, \frac{a}{\sqrt{2(1-\lambda)}}$. This type of spin configuration defines a `meronic' type of the spin vortex.
\paragraph*{}
The definition of the spin-spin correlation function is given by,
\begin{eqnarray}
S(\mathbf{r},t)&=& \langle \mathbf{S}(\mathbf{r},t)\cdot \mathbf{S}(\mathbf{0},0)\rangle \nonumber \\
&=& \langle S^{x}(\mathbf{r},t)S^{x}(\mathbf{0},0)\rangle +\langle S^{y}(\mathbf{r},t)S^{y}(\mathbf{0},0)\rangle + \langle S^{z}(\mathbf{r},t)S^{z}(\mathbf{0},0)\rangle ,
\end{eqnarray}
where $\langle ... \rangle$ represents the thermal average. In the case of classical gas of ideal vortices the thermal average has to be done by taking Maxwellian velocity distribution function. Here $S^{xx}(\mathbf{r},t)= \langle S_{x}(\mathbf{r},t)S_{x}(\mathbf{0},0)\rangle$ and $S^{yy}(\mathbf{r},t)\,=\, \langle S_{y}(\mathbf{r},t)S_{y}(\mathbf{0},0)\rangle$ are in-plane correlations and $S^{zz}(\mathbf{r},t)\, \,= \,\langle S_{z}(\mathbf{r},t) \\ S_{z}(\mathbf{0},0)\rangle$ is the out-of-plane correlation.
\paragraph{}
The effective analytical expression for the in-plane correlation can be taken as \cite{14},
\begin{equation}
S^{xx}(\mathbf{r},t)= \frac{S^2}{2}exp\lbrace\left[\frac{r^2}{\xi^2}+\gamma ^2 t^2\right]^{1/2}\rbrace ,
\end{equation}
with $\gamma= \frac{\sqrt{\pi \bar{u}}}{2\xi}$, where $\bar{u}$ is the root mean square velocity. Here $\xi = \xi_{0}e^{b/\sqrt{\tau}}$ is the vortex- vortex correlation length. The root-mean squared velocity of the vortices was first calculated by Huber as \cite{20};
\begin{equation}
\bar{u}\,=\,\sqrt{b\pi}\frac{JS(S+1)a^2}{\hbar}\sqrt{n_{v}^{f}}\tau^{-1/4},
\end{equation}
where $n_{v}^{f}$ is the density of free vortices at $T > T_{KT}$. The Fourier transform of  $S^{xx}(\mathbf{r}, t)$ in (5) gives rise to the in-plane dynamical structure function given by,
\begin{equation}
S^{xx}(\mathbf{q},\omega)= \frac{S^2}{2\pi ^2}\frac{\gamma ^3 \xi ^2}{[\omega ^2 + \gamma ^2(1+\xi ^2 q^2)]^2}.
\end{equation}
This is a squared Lorentzian, peaked at $\omega = 0$, with q dependent width,
\begin{equation}
\Gamma = \frac{1}{2}\lbrace \pi(\sqrt{2}-1)\rbrace ^{1/2}\left(\frac{\bar{u}}{\xi}\sqrt{1+\xi^2 q^2}\right).
\end{equation}
Exactly same results holds for $S^{yy}(\mathbf{q},\omega)$ also.
\paragraph*{}
From the definition of $S^{zz}(\mathbf{r},t)$, it can be shown that the out-of-plane correlation is given by \cite{14},
\begin{eqnarray}
S^{zz}(\mathbf{r},t) &=& n_{v}^{f} S^2 \int\int d^{2}R\,\,d^{2}u P(\mathbf{u}) \cos\theta(\mathbf{r}-\mathbf{R}-\mathbf{u}t)\cos\theta(\mathbf{R}),
\end{eqnarray} 
where $P(\mathbf{u})$ is the Maxwell velocity distribution for a single vortex. Performing first the spatial Fourier transform and then the temporal Fourier transform, it can be shown that, the out-of-plane dynamical structure function has the form,
\begin{equation}
S^{zz}(\mathbf{q},\omega)= \frac{S^2}{4\pi^{5/2}}n_{v}^{f}\frac{|f(q)|^2}{\bar{u}q}exp\left(- \frac{\omega^2}{\bar{u}^2 q^2}\right).
\end{equation}
Here $|f(q)|$ is the velocity independent vortex form factor and it has the form $f(\mathbf{q})=\int d^{2}\mathbf{r}\cos\theta(\mathbf{r})e^{-i\mathbf{q}\cdot\mathbf{r}}$. The form of $S^{zz}(\mathbf{q},\omega)$ as in $eq^n$(10) exhibits a central peak at $\omega\,\,=\,\,0$. The width of the central peak is $\Gamma_{z}\,\,=\,\, \bar{u}q$ i.e., linear in $q$.
\paragraph*{}
In a typical inelastic neutron scattering experiment the count rate is related to the dynamical structure function as  \cite{23},
\begin{equation}
I(\mathbf{q},\, \omega) \propto \,\,S(\mathbf{q}, \omega).
\end{equation}
At finite temperature there always exist creation and annihilation of excitations. A detailed balance condition is always needed to relate the intensities of up scattering ($\hbar \omega < 0$) and down scattering ($\hbar \omega > 0$). True quantum mechanical $S(\mathbf{q},\omega)$, denoted by $S_{DB}(\mathbf{q},\omega)$ is recovered by the relation ,
\begin{equation}
S_{DB}(\mathbf{q},\omega)\,=\, \frac{2}{1+exp(\frac{-\hbar \omega}{k_{B}T})}S(\mathbf{q},\omega).
\end{equation}
Where the factor $\frac{2}{1+exp(\frac{-\hbar \omega}{k_{B}T})}$ is called the Windsor factor  \cite{17}. This $S_{DB}(\mathbf{q},\omega)$ incorporates the detailed balance condition, as required by the thermal equillibrium. Another important factor, which has to be taken into account, is the instrumental resolution function $R(t)$ or $R(\omega-\omega^{\prime})$. This essentially incorporates the different independent instrumental properties that affect the incident and scattered beam of neutrons \cite{23}. In order to compare theory with experiment one has to convolute the theoretical expression, obtained from a model under consideration, by the resolution function \cite{24}. Thus we consider the convoluted dynamical structure function $S_{conv.}(\mathbf{q},\, \omega)$ given by,
\begin{eqnarray}
S_{conv.}(\mathbf{q},\, \omega) &=& \int dt \int d^2 r R(t) S(\mathbf{r},t) e^{i(\mathbf{q}\cdot\mathbf{r}- \omega t)} \nonumber \\ 
&=& \int R(\mathbf{q}, \omega-\omega^{\prime})S(\mathbf{q}, \omega^{\prime})d\omega^{\prime}.
\end{eqnarray}
\paragraph*{}
The resolution function has to be chosen so as to give minimum ripples at the end points of the resolution width. For this purpose a suitable window function such as Tukey window function may be chosen,
\begin{eqnarray}
R(t)&=& \frac{1}{2}\left[ 1+\cos (2\pi t/t_{m})\right] \,\, for\,\, |t|\leq t_{m}/2\nonumber\\
&=& 0 \,\,otherwise
\end{eqnarray}
The parameter $t_{m}$, occurring in the Window function, can be set from the resolution half width obtained from experimental data.
\subsection{In-Plane Dynamical Structure Function}
In our formulation for the in-plane dynamical structure function we take into account the Tukey window function, as mentioned above (see equations (14)). Using (5), (13) and (14) we compute the Fourier transform of in-plane spin-spin correlation,
\begin{eqnarray}
S^{xx}_{conv.}(\mathbf{q},\omega) &=& \frac{1}{(2\pi)^{3/2}} \int d^{2}r \int_{\frac{-t_m}{2}}^{\frac{t_m}{2}} dt \,\, S^{xx}(\mathbf{r},t) R(t) \times  \nonumber \\ &\qquad{}& e^{i(\mathbf{q}\cdot\mathbf{r}- \omega t)}.
\end{eqnarray}
Now, $\int d^2 r e^{i\mathbf{q}\cdot\mathbf{r}}= \int_{0}^{\infty} r\,\,dr \int_{0}^{2\pi}\,\, d \theta e^{iqr\, cos\theta}= \int_{0}^{\infty} r\,\,dr\,\,J_{0}(qr)$, where $J_{0}(qr)$ is Bessel function of order zero. The spatial integration is performed from zero to a certain radius $R_0$. 
A final expression for the convoluted in-plane dynamical structure function takes the form,
\begin{eqnarray}
S^{xx}_{conv}(\mathbf{q},\omega)&=& \frac{1}{(2\pi)^{1/2}} \int_{0}^{R_0} dr \int_{\frac{-t_m}{2}}^{\frac{t_m}{2}} dt \,\, S^{xx}(\mathbf{r},t)\times \nonumber \\
&\qquad{}& r J_{0}(qr) R(t) cos(\omega t).
\end{eqnarray}
Since, $S^{xx}(\mathbf{r},t)$ and $R(t)$ are both even function in $t$ , only $cos(\omega t) $ contributes to the temporal part of the integration. From symmetry Y component of the in-plane dynamical structure function $S^{yy}_{conv}(\mathbf{q},\omega)$ is same as X component of the in-plane dynamical structure function $S^{xx}_{conv}(\mathbf{q},\omega)$. Let us note that in the above analysis the formulation holds only for $T > T_{KT}$. For $T < T_{KT}$ the vortex- vortex correlation length $\xi$ is not defined and hence the formalism can't be extrapolated below $T_{KT}$.
\subsection{Out-of-Plane Dynamical Structure Function}
The out-of-plane dynamical structure function is given by,
\begin{equation}
S^{zz}_{conv}(\mathbf{q},\, \omega)=\,\int R(\omega-\omega^{\prime})S^{zz}(\mathbf{q}, \omega^{\prime})d\omega^{\prime},
\end{equation}
where $R(\omega-\omega^{\prime})$ is the Fourier transform of R(t). The reason for taking (17) as the expression for convoluted out-of-plane dynamical structure function is that, unlike (5), an analytical expression for $S^{zz}(\mathbf{r}, \, t)$ can't be evaluated from (9). So one has to start from $eq^n$(10).
\paragraph*{}
The integral in (17) has been computed numerically. The $S^{zz}_{conv}(\mathbf{q},\, \omega)$ defined here corresponds only to the mobile vortices; whereas the experimental data contain the contributions from bound vortices and fragile `spin wave like' modes. These fragile `spin wave like' modes are the largely decaying spin wave modes above the ferromagnetic- paramagnetic transition temperature (Curie temperature). In order to compare with the experimental observations, one has to extract the mobile vortex contribution from the experimental data. This can be done by subtracting the fragile mode contribution and the frozen vortex contribution from the experimental data. The fragile mode contribution has been subtracted by taking the fragile mode contribution above transition temperature to be same as the spin wave contribution just below transition temperature. This is valid as long as we are considering the temperature which are not far below or above from the transition temperature.
To find the approximate analytical expression for $S^{zz}(\mathbf{q}, \omega)$ due to bound vortex contribution, the limiting value of $\bar{u}$ is taken as $\bar{u}\rightarrow0$ in (10). Then from (10) it easy to find an expression for $S^{zz}_{bound}(\mathbf{q},\omega)$ namely,
\begin{equation}
S^{zz}_{bound}(\mathbf{q},\omega)= \frac{S^2}{4\pi^{2}}n_{v}^{b}|f(q)|^2 \delta(\omega),
\end{equation}
Where $n_{v}^{b}$ is the bound vortex density. Since, the system has no net topological charge we can assume that there are equal number of vortices and anti-vortices present in the system and we can take $n_{v}^{f}  + n_{v}^{b}= \frac{1}{2} $ assuming square lattice structure. This is correct as long as the temperature is just below $T_{KT}$ where all the vortices are frozen but once the temperature crosses $T_{KT}$ some of the bound vortices become mobile and the bound vortex density can be approximated as,
\begin{equation}
n_{v}^{b}\approx(\frac{1}{2} - n_{v}^{f}),
\end{equation}
where, $n_{v}^{b}$ is in the units of inverse of plaquette size($a^2$). 
Since, $n_{v}^{f}\sim\xi_{0}^{-2}exp(-2b/\sqrt{\tau})$ \cite{14}, $n_{v}^{b}$ given by (19) is temperature dependent. Here $\xi_{0}$ is of the order of lattice parameter. Using (18) and (19) the bound vortex contribution has to be subtracted carefully from the experimentally observed count.
\paragraph*{}
We would like to point out that we could not apply this procedure to extract out bound vortex contributions in the case of in-plane dynamical structure function (see Sec: 3).
\subsection{Total Dynamical Structure Function (Spin-Spin Correlation)}
The general expression for the total dynamical structure function is,
\begin{equation}
S(\mathbf{q},\omega)= \frac{1}{(2\pi)^{3/2}} \int d^{2}r \int dt \,\, S(\mathbf{r},t) R(t) e^{i(\mathbf{q}\cdot\mathbf{r}- \omega t)}.
\end{equation}
where, the total spin-spin correlation is$S(\mathbf{r},t)$ is defined by (4).  So, the total dynamical structure function is $S(\mathbf{q},\omega)=S^{xx}(\mathbf{q},\omega)+S^{yy}(\mathbf{q},\omega)+S^{zz}(\mathbf{q},\omega)$. Since, X and Y components of the spins are symmetric we have, $S^{xx}(\mathbf{q},\omega)=S^{yy}(\mathbf{q},\omega)$and the total dynamical structure function takes the form,
\begin{equation}
S(\mathbf{q},\omega)=2S^{xx}(\mathbf{q},\omega)+S^{zz}(\mathbf{q},\omega).
\end{equation}
Here, we would consider (21) only for mobile vortices.
\paragraph*{}
It is an important fact that the formalism explained above incorporates the Windsor factor and the presence of $\hbar$ in the quantum expression of magnetic moment corresponding to the spins constituting the vortex \cite{13}. Therefore the formalism looks like a semi-classical one. Henceforth we will call our combined theoretical approach `semi-classical like'.
\section{Calculations and Results}
We apply the formalism of Sec: 2 on a real material $K_2 Cu F_4$ for which neutron scattering experiments have been performed \cite{6}. It is a quasi-two-dimensional spin-$\frac{1}{2}$ ferromagnet, where the interaction is mainly Heisenberg type with only 1$\%$ X-Y like anisotropy . The transition is close to KT type with slight modification due to Heisenberg type interaction. The magnetic lattice structure for $K_2CuF_4$ is approximately a body centred tetragonal lattice i.e. a lattice, composed of stacking of 2D square lattices \cite{6}. The physical parameters are given in the Table 1, which have been used throughout  the calculation.
\begin{table}
\caption{Relevant parameters for $K_2CuF_4$ \cite{6}}
\label{tab:1}       
\centering
\begin{tabular}{p{4.5cm}l}
\hline\noalign{\smallskip}
parameter & magnitude \\
\noalign{\smallskip}\hline\noalign{\smallskip}
exchange coupling (J) & 11.93 K  \\
lattice parameter ($a$) & 4.123 \AA \\
`b' & 1.5 \\
`$T_{KT}$' & 5.5 K \\
\noalign{\smallskip}\hline
\end{tabular}
\vspace*{0.5cm}  
\end{table}
\paragraph{}
We start with the investigation of the in-plane correlation (in-plane dynamical structure function) $S^{xx}(\mathbf{q}, \omega)$. The radius $R_0$ in (16) is $(\sqrt{100^2 + 100^2})a$ for a $100 \times 100$ lattice, as used in the MCMD analysis by Mertens et. al. \cite{14}, where $a$ is the lattice parameter.  We set the value of $t_m$ according to the experimental resolution width (0.01 meV) \cite{6}. We compute $S^{xx}_{conv}(\mathbf{q}, \omega)$ numerically, for two different temperatures, 6.25 K and 6.75 K, for q(planar)= 0.04 reciprocal lattice units(in the units of $\frac{\pi}{a}$), experimentally $\hbar q$ being the momentum transfer.
\begin{figure}
\centering
\resizebox{0.65\textwidth}{!}{%
  \includegraphics{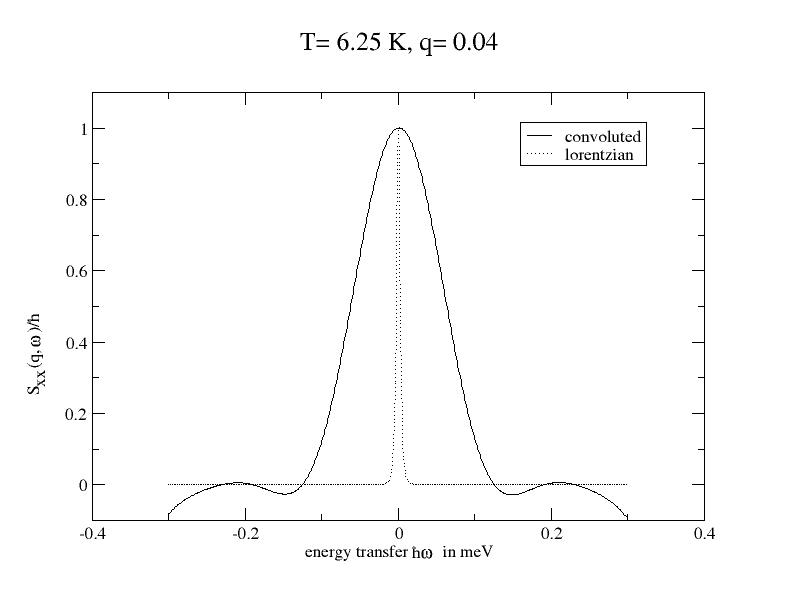}
}
\vspace{0.5cm}
\caption{Comparison between the convoluted in-plane dynamical structure function $S_{conv}^{xx}(\mathbf{q}, \omega)$ (eqn. 16) and unconvoluted in-plane dynamical structure function $S^{xx}(\mathbf{q}, \omega)$ (eqn. 7)at  T = 6.25 K and q = 0.04. Solid line is for convoluted theoretical expression and dotted line is for unconvoluted theoretical expression(squared Lorentzian). $\xi= 58.09 a$,  $\bar{u}= 0.0614 \frac{a}{t_{nat}}$, and width $\Gamma_{xx}= 0.0012 meV$ for squared Lorentzian at T= 6.25 K.}
\label{Figure 1}
\end{figure}
\begin{figure}
\centering
\resizebox{0.65\textwidth}{!}{%
  \includegraphics{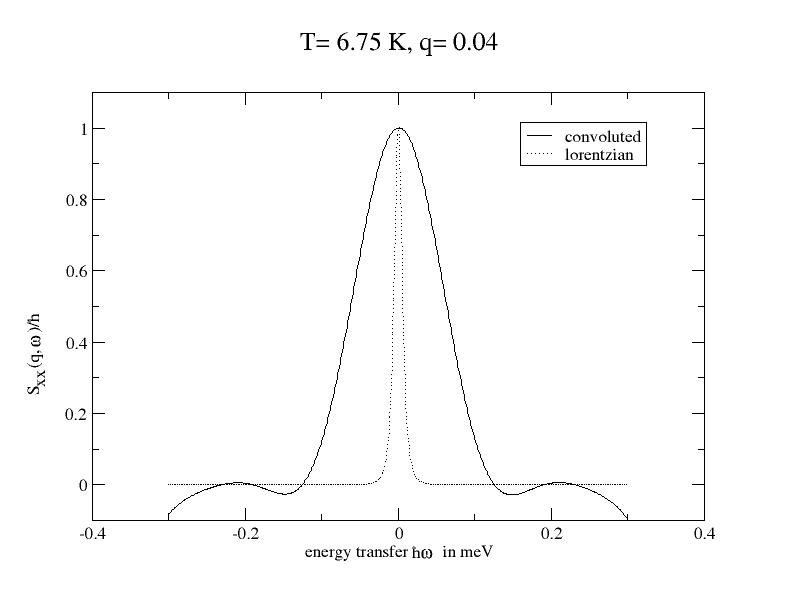}
}
\vspace{0.5cm}
\caption{Comparison between the convoluted in-plane dynamical structure function $S_{conv}^{xx}(\mathbf{q}, \omega)$ (eqn. 16) and unconvoluted in-plane dynamical structure function $S^{xx}(\mathbf{q}, \omega)$ (eqn. 7)at  T = 6.75 K and q = 0.04. Solid line is for convoluted theoretical expression and dotted line is for unconvoluted theoretical expression(squared Lorentzian). $\xi= 22.25 a$,  $\bar{u}= 0.1352 \frac{a}{t_{nat}}$ and width $\Gamma_{xx}= 0.0035 meV$ for squared Lorentzian at T= 6.25 K.}
\label{Figure 2}
\end{figure}
There are two threshold values of $q$ \cite{6}, namely $q_1 = 0.06$ and $q_2 = 0.01$, where for $q>q_1$ the system behaves like 2D Heisenberg system and  for $q<q_2$ the system behaves as 3D XY system. For $q_2 < q < q_1$ the system behaves as 2D XY system.We have varied the energy transfer $\hbar\omega$ from -0.3 meV to +0.3 meV, which includes the range -0.2 meV to +0.2 meV as taken in experiment \cite{6}. The convoluted in-plane dynamical structure function is plotted in Fig. 1 and Fig. 2, where $t_{nat} = \frac{\hbar}{JS(S+1)}$ is the natural time unit for the system/material (in our case $K_{2}CuF_{4}$). These figures indicate that after convoluting with the Tukey window function, the in-plane dynamical structure function no longer remains squared Lorentzian, though in both the cases central peaks persist. The width of the $S^{xx}_{conv}(\mathbf{q}, \omega)$ curve is much larger than that of the squared Lorentzian. 
\paragraph*{}
We notice that the convoluted in-plane dynamical structure function function $S^{xx}_{conv}(\mathbf{q}, \omega)$ has become negative just above 0.1 meV. The occurence of negative values of the dynamical structure function has been dealt with in detail in sec.-4 and in Appendix.
\paragraph*{}
Again comparing Fig. 1 and Fig. 2 we find that the width of the squared Lorentzian increases with the increase of temperature whereas that of the $S^{xx}_{conv}(\mathbf{q}, \omega)$ does not undergo any change. Later, we will present a comparison of the convoluted total dynamical structure function with the experimental one(See Fig. 5 and Fig. 6).
\paragraph*{}
We now evaluate the out-of-plane dynamical structure function $S^{zz}_{conv}(\mathbf{q}, \omega)$ for two different temperatures, 6.25 K and 6.75 K, for q(planar ) = 0.04 r.l.u, using (17). The expression for $R(\omega - \omega^{\prime})$ is,
\begin{equation}
R(\omega - \omega^{\prime})= \frac{1}{4\pi}sin \left[ \frac{(\omega-\omega^{\prime})t_m}{2} \right] [  \frac{2}{\omega  - \omega^{\prime}} \frac{1}{\omega-\omega^{\prime}+2\pi /t_m}- \frac{1}{\omega-\omega^{\prime}-2\pi /t_m} ] . 
\end{equation}
We use the same value of $t_m$ as used for $S^{xx}_{conv}(\mathbf{q}, \omega)$. Here also the reasons for the choice of temperatures and q(planar) are same as that for the in-plane dynamical structure function. In Fig. 3 and Fig. 4 we have plotted the out-of-plane correlation, $S^{zz}_{conv}(\mathbf{q}, \omega)$. We have varied the $\omega^{\prime}$ from $-\frac{\pi}{t_m}$ to $\frac{\pi}{t_m}$ in (17), where $t_m$ is estimated from the resolution width as before.
\begin{figure}
\centering
\resizebox{0.65\textwidth}{!}{%
  \includegraphics{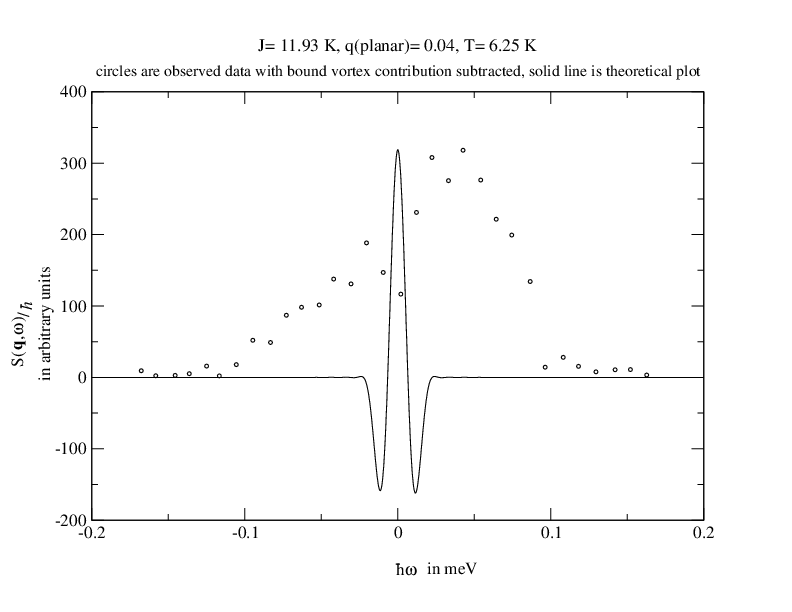}
}
\vspace{0.6cm}
\caption{circles are observed (experimental) data, where contributions from the fragile modes as well as the bound vortex contributions have been subtracted \& solid line is the plot of properly convoluted out-of-plane dynamical structure function $S^{zz}(\mathbf{q}, \omega)$(theoretical). $\xi= 58.09 a$,  $\bar{u}= 0.0614 \frac{a}{t_{nat}}$.}
\label{Figure 3}
\end{figure}
\begin{figure}
\centering
\resizebox{0.65\textwidth}{!}{%
  \includegraphics{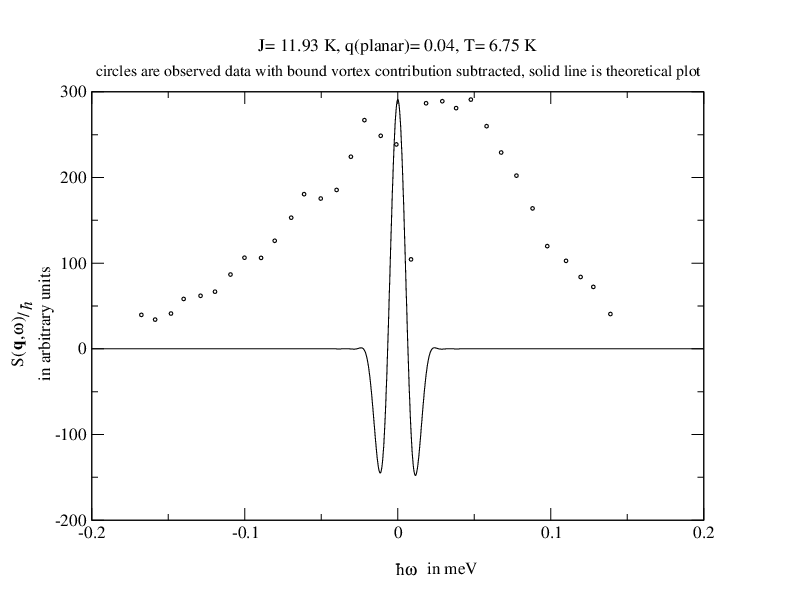}
}
\vspace{0.6cm}
\caption{circles are observed (experimental) data, where only contributions from bound vortices have been subtracted \& solid line is the plot of properly convoluted out-of-plane dynamical structure function $S^{zz}(\mathbf{q}, \omega)$ (theoretical). $\xi= 22.25 a$,  $\bar{u}= 0.1352 \frac{a}{t_{nat}}$.}
\label{Figure 4}
\end{figure}
\paragraph*{}
In this case the bound vortex contribution has been subtracted carefully, using (18) and (19), from the observed count at 6.25 K to obtain the effective mobile vortex contribution. The methodology for extracting the mobile vortex contributions from the experimental data has been explained in Sec 2.2. As long as the counts at 6.75 K are concerned, the fragile `spin wave like' modes are highly decaying so that it can't be assumed to be the same as the true spin wave modes observed at 5 K. So only bound vortex contribution has been subtracted at 6.75 K. The normalization factors, required for the quantitative comparison between the theoretical and the experimental results, have been estimated from the neutron count extracted from the experiment on $K_{2}CuF_{4}$  \cite{6}.
\paragraph*{}
We find that the out-of-plane dynamical structure function is also negative within the resolution width(see Figs. 3 and 4)! Moreover experimental peak is out side the resolution width, while the peak corresponding to the $S^{zz}_{conv}(\mathbf{q}, \omega)$ is at $\omega = 0$.
\paragraph*{}
The above calculations lead us to the theoretical estimate for the convoluted total dynamical structure function $S^{total}_{conv}(\mathbf{q}, \omega)$ given by (21). In Figs. 5 and 6, $S^{total}_{conv}(\mathbf{q},\omega)$ has been compared with the filtered experimental data obtained by subtracting the bound vortex contributions and fragile `spin wave like' contributions (see Sec. 2.1 and 2.2). In these plots the intensities of the experimental peak and that of the central peak of the $S^{total}_{conv}(\mathbf{q},\omega)$ have been matched. 
\begin{figure}
\centering
\resizebox{0.65\textwidth}{!}{%
  \includegraphics{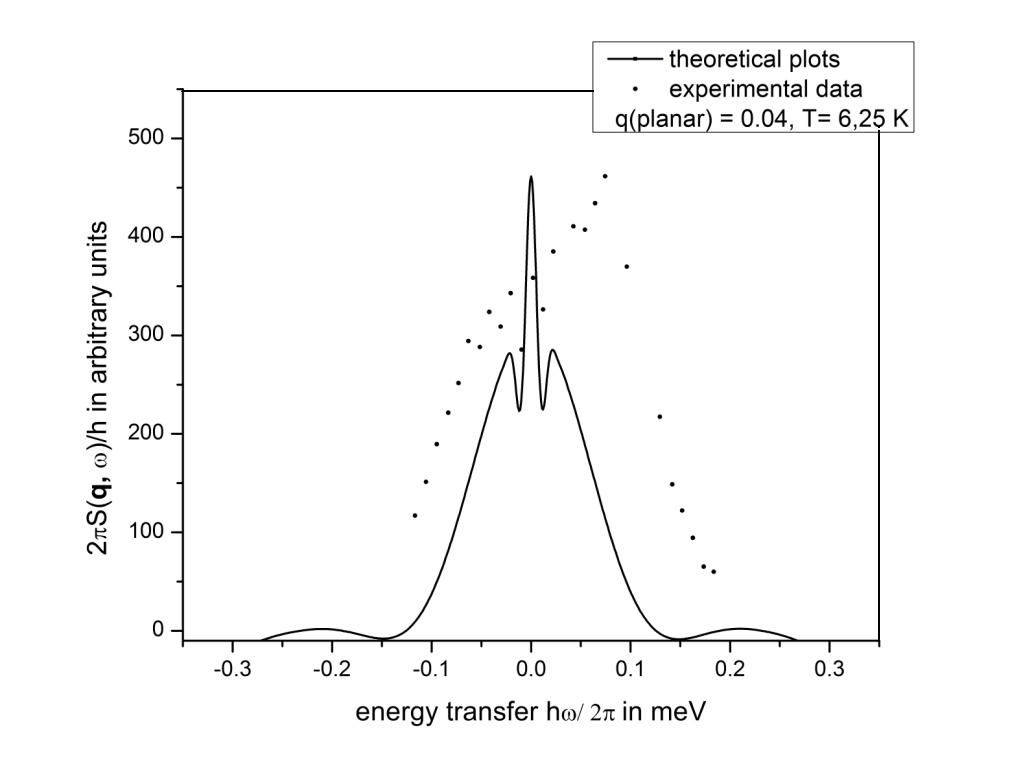}
}
\vspace{0.5cm}
\caption{total dynamical structure function $S^{total}_{conv}(\mathbf{q}, \omega)$ at  T = 6.25 K and q = 0.04- solid line is for convoluted theoretical results and dots are filtered experimental data. $\xi= 58.09 a$,  $\bar{u}= 0.0614 \frac{a}{t_{nat}}$.}
\label{Figure 5}
\end{figure}
\paragraph*{}
It is clear from Fig. 5 that at 6.25 K the experimental peak occurs approximately at 0.08 meV, which is way outside the resolution width. At 6.75 K [see Fig. 6] the peak of the experimental graph is not far from the central peak. It is reasonable to say that as the temperature is increased, we are getting better agreement of the $S^{total}_{conv}(\mathbf{q},\omega)$ with the experimental observations. This agreement isregarding the position of the central peak. Apart from the central peak there are two other peaks at finite frequency at both the temperatures. These are nothing but the reminiscent of the out-of-plane dynamical structure function contribution as seen from Figs. 3, 4, 5 and 6. This signifies the fact that the in-plane correlation is largely dominating over the out-of-plane correlation. 
\paragraph*{}
The total spin-spin correlation is still negative just above 0.1 meV. 
\begin{figure}
\centering
\resizebox{0.65\textwidth}{!}{%
  \includegraphics{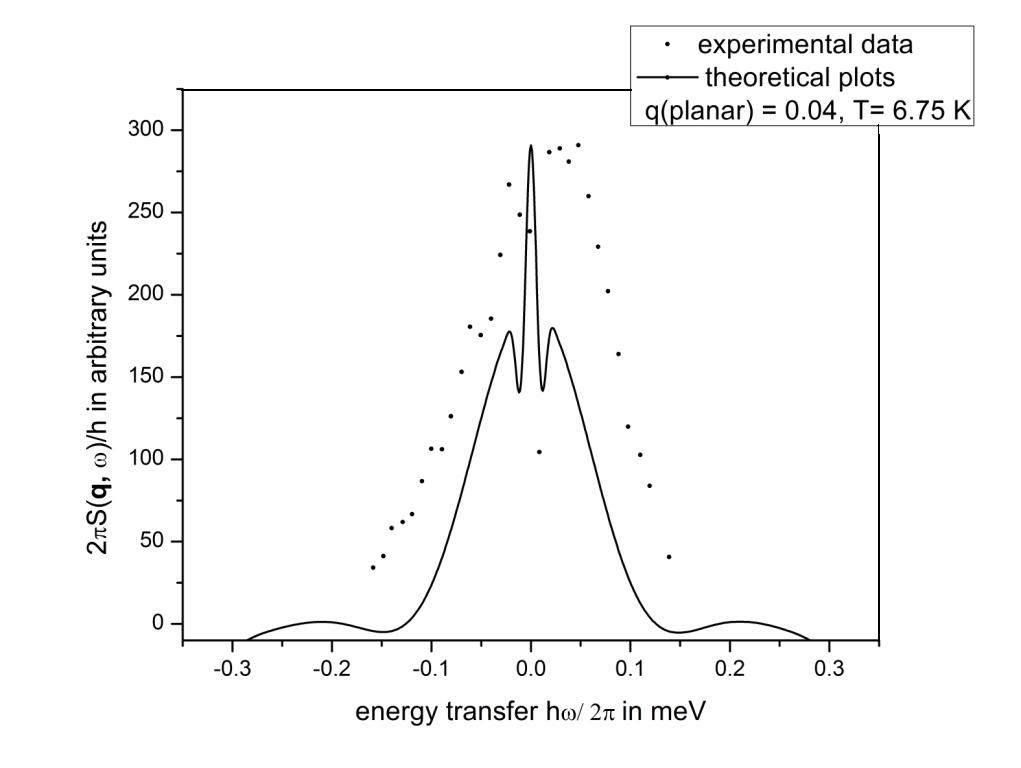}
}
\vspace{0.5cm}
\caption{total dynamical structure function $S^{total}_{conv}(\mathbf{q}, \omega)$ at  T = 6.75 K and q = 0.04- solid line is for convoluted theoretical results and dots are filtered experimental data. $\xi= 22.25 a$,  $\bar{u}= 0.1352 \frac{a}{t_{nat}}$.}
\label{Figure 6}
\end{figure}
Though it is true that the dynamical structure function can't be negative, here in our case the negativity occurs as a result of the convolution of analytical expression of $S(\mathbf{q}, \omega)$. Even for a conventional long range ordered system, the dynamical structure function corresponding to a classical pure spin wave comes out to be negative beyond a certain range of frequency when convoluted with any spectral window function. Furthermore the above peculerity persists even when quantum effects are incorporated through a detailed balance factor[\textbf{see Appendix}].
\paragraph*{}
The inclusion of quantum mechanical detailed balance factor in the semi-classical like treatment for dynamics of mobile vortices and anti-vortices, is not even causing any appreciable asymmetry, as seen in the theoretical plots in our case of spin-$\frac{1}{2}$. The theoretical plots are largely symmetric around $\omega = 0$. A very small asymmetry in the theoretical plots are being seen for higher values of $\omega$ while the experimental data are showing clearly the asymmetry.
\paragraph*{}
It may be noted that in our analysis the bound vortex contributions have been approximately estimated only for out-of-plane dynamical structure function $S^{zz}(\mathbf{q}, \omega)$. This is because, in this case, we are able to truncate the expression, as given in (10), to the regime $T<T_{KT}$, by making $\bar{u} \rightarrow 0$. In (10), there exists no explicit dependence of $S^{zz}(\mathbf{q}, \omega)$ on the correlation length $\xi$. In case of in-plane correlation, as given in (5), we need to find $\xi$ for $T<T_{KT}$ due to its explicit appearance in that expression. Since $\xi$ is not defined for $T<T_{KT}$ we are not able to estimate the bound vortex contribution for in-plane dynamical structure function. 
\paragraph*{}
In summary, we find that the width of the convoluted in-plane dynamical structure function is much larger than that of the squared Lorentzian. Values of the in-plane dynamical structure function comes out to be negative beyond a finite range of energy transfer. The convoluted out-of-plane dynamical structure function becomes negative as well; however this happens within the resolution width about the central peak (peak at $\omega = 0$). The total convoluted dynamical structure function also  becomes negative in the regime where the in-plane dynamical structure function had become negative. No appreciable asymmetry is created even after including the Windsor factor. We find that for both the temperatures the convoluted total dynamical structure function is symmetric around $ \omega = 0$; whereas the experimental observation is not. The theoretical model of semi-classical treatment of ideal gas of unbound vortices tends to agree with the experimental observations better at higher temperatures (for spin-$\frac{1}{2}$ system), when we consider the experimental results at T= 6.25 K and T= 6.75 K (Figs. 5 and 6). It is worthwhile to point out that same results hold for unbound anti-vortices also.  
\section{Conclusions \& Discussions}
The laws of quantum mechanics which govern all real systems, ensure the dynamical structure functions to be always positive definite \cite{23}. We find in our analysis however, that the semi-classical treatment based on ideal gas of vortices (anti-vortices) for a low spin system leads to the occurence of negative values of dynamical structure function, over a large range of energy transfer, when convoluted with any standerd resolution function.
\paragraph*{}
Based on the analysis carried out in the \textbf{Appendix} we can infer that for the dynamics of mobile vortices and anti-vortices, the negative values of $S_{conv.}(\mathbf{q}, \omega)$ are occurring due to the following factors; \\
i) the choice of the resolution function, whicn in our case is the Tukey function,\\ ii) the choice of the value of resolution width $\Delta\omega$, which in this case is made fixed by experimentally imposed resolution width, \\ iii) use of a semi-classical like treatment to extend the classical theory of dynamics of mobile vortices (anti-vortices) to a quasi two dimensional spin $\frac{1}{2}$ ferromagnet which is quantum mechanical.
\paragraph*{}
To avoid the negativity in the $S_{conv.}(\mathbf{q}, \omega)$ we could have chosen a different resolution function. Indeed, it has been shown that most of the resolution functions are more or less oscillatory in the fourier space \cite{24,29} . An extra smoothening factor can be used to dampen the oscillation of the resolution function. This extra factor is eventually related to the resolution width and it makes the resolution function smoother if the resolution width is decreased \cite{24}. However, in our case the resolution width is fixed from the experiment and consequently the oscillation of the resolution function can't be avoided by merely changing the resolution function.
\paragraph*{}
It has been found that there is a range of $\omega$ over which $S_{conv.}(\mathbf{q}, \omega)$ remains positive. We can call it as the physically admissible range. This range is related to the magnitude of the spin occuring in the theoretical model under consideration and to the resolution width. On the basis of the analysis (presented in the \textbf{Appendix}) it is expected that even in the case of dynamics of mobile vortices and anti-vortices the physically admissible range would be larger for higher spin value and smaller for lower spin values.
\paragraph*{}
Another way to avoid the negativity is to assume, $S_{conv.}(\mathbf{q}, \omega) = 0 $ outside the physically admissible range \cite{24}. If this physically admissible range is within the range of experimental interest then the assumption is not applicable. In our case of spin-half ferromagnet the physically admissible range is well within the range of $\omega$ over which the neutron scattering data has been taken in the experiment (as seen from Figs. 5 and 6).
\paragraph*{}
Hence, the negative values of $S_{conv.}(\mathbf{q}, \omega)$ can only be due to the use of the semi-classical like treatment to extend the classical theory of dynamics of mobile vortices and anti-vortices to a quasi two dimensional spin $\frac{1}{2}$ ferromagnet ($K_2 CuF_4$).
\paragraph*{}
Moreover, the convoluted out-of-plane dynamical structure function computed from our semi-classical like treatment, becomes negative within the experimentally imposed resolution width itself. Thus the central peak occurring in this case may not posses a well defined width.
\paragraph*{}
The agreement between the behaviour of dynamical structure functions obtained from our theoretical calculations and that from the experiment, in terms of the peak position and the overall shape, is found to be fairly good at temperatures much larger than $T_{KT}$.
\paragraph*{}
Although the vortices (and anti-vortices) are extended objects the Maxwell- Boltzmann distribution can still be used for the motion of the centre of mass of these objects.
\paragraph*{}
Our investigation presented in this paper brings out the fact that a complete quantum mechanical treatment is essential for describing the detailed features of the dynamics of unbound spin vortices and anti-vortices corresponding to low spin magnetism systems. As a first step towards this, a theoretical framework for describing static quantum spin vortices and anti-vortices and their topological properties has been developed \cite{25,26,27,28}. An extension of this formalism to the case of mobile spin vortices and anti-vortices is crucial for the quantum mechanical calculation of dynamical structure function. This would go a long way towards an explanation for the experimental results observed for the genuine quantum spin systems like $K_2 CuF_4$.   

\section{Acknowledgments}
One of the authors (SS) acknowledges the financial support through Junior Research Fellowship (09/ 575 (0089) / 2010 EMR--1) provided by Council of Scientific \& Industrial Research (CSIR) and also acknowledges Mr. Bandan Chakrabortty for valuable discussions regarding the computer programming.

\section*{Appendix: $S_{conv}(\mathbf{q},\, \omega)$ corresponding to classical spin-wave}
For classical spin wave at very low temperature corresponding to a classical Heisenberg ferromagnet, the dynamical structure function has the form:-
\begin{eqnarray}
S(\mathbf{q},\, \omega) &=& \delta(\omega^{2} - \omega_{q}^{2}) \nonumber \\ 
 &=& \frac{1}{2\omega_{q}}[\delta(\omega - \omega_{q}) + \delta(\omega + \omega_{q})].
\end{eqnarray}
Where, $\omega_{q} = \hbar^{-1}J\sqrt{S(S+1)} \,\,z (1-\gamma_{q})$ for a cubic lattice.
Here $z$ is the number of nearest neighbours and $\gamma_{q} = \frac{1}{z} \sum_{r} cos(\mathbf{q} \cdot \mathbf{r})$.
\paragraph*{}
Convoluted dynamical structure function, $S_{conv}(\mathbf{q},\, \omega)$ has been defined in (13), where $R(\omega - \omega^{\prime})$ is the fourier transform of a siutably chosen spectral function. So far Tukey function has been used and the same is used in this case also. The fourier transform of Tukey function is given by (22). Using (13), (22) and (23) we find the convoluted dynamical structure function corresponding to spin wave, $S^{SW}_{conv}(\mathbf{q},\, \omega)$ to be,
\begin{equation}
S^{SW}_{conv}(\mathbf{q},\, \omega) = \frac{1}{2\omega_{q}}[R(\omega - \omega_{q}) + R(\omega + \omega_{q})].
\end{equation}
Analyzing (24) we find that there exists a range 
\begin{equation}
|\omega| \leq (\omega_{q} + \frac{4\pi}{t_{m}})
\end{equation}
within which $S^{SW}_{conv}(\mathbf{q},\, \omega)$ remains positive and outside, it becomes negative. This remains true even if we choose any other resolution function. This is basically due to the fact that when fourier transform is performed on spectral functions (defined in time domain), the resulting functions in $(\mathbf{q}, \omega)$ space mostly turn out to be oscillatory \cite{29}.
\paragraph*{}
One way to avoid these negative values of $S_{conv}(\mathbf{q},\, \omega)$ would be to assume $S_{conv}(\mathbf{q},\, \omega) = 0$ outside $|\omega| \leq (\omega_{q} + \frac{4\pi}{t_{m}})$ \cite{24}. This prescription however, can't be taken into consideration when comparing theoretical predictions with experimental results, if the energy range of experimental interest contains the above mentioned range of $\omega$. Another way to avoid the negative values of $S_{conv}(\mathbf{q},\, \omega)$ is to decrease the value of $t_{m}$ [see (25)], where $t_{m}$ is related to the experimental resolution width ($\Delta \omega$) by the relation,
\begin{equation}
t_{m} \simeq \frac{\hbar}{2\Delta \omega},
\end{equation} 
$\Delta\omega$ being in energy units. Decrease in $t_{m}$ means increase in the value of $\Delta\omega$ which signifies a poor experimental resolution width. In comparing the theoretical predictions with the experimental results, $t_{m}$ is determined by the relation (26), where $\Delta \omega$ is fixed from the experiment.
\paragraph*{}
Another interesting feature of (25) is the fact that the region of $\omega$ where $S_{conv}(\mathbf{q},\, \omega)$ remains positive, depends on the spin value S via the relation,
\begin{equation}
|\omega| \leq (\hbar^{-1}J\sqrt{S(S+1)} \,\,z [1-\gamma_{q}] + \frac{4\pi}{t_{m}}).
\end{equation}
This shows that when $t_{m}$ is fixed from the experimental resolution width, S remains as the free parameter to determine the range of $\omega$ as mentioned in (25). Increase in S will increase the range of $\omega$ where $S_{conv}(\mathbf{q},\, \omega)$ remains positive. Hence for a system with high spin value S, the range of positive $S_{conv}(\mathbf{q},\, \omega)$ may actually be at par with the energy range of experimental interest, unlike the case of spin $\frac{1}{2}$.
\paragraph*{}
It can be shown from (24) that the range of $\omega$, given by (25), will not change even if a detailed balance factor (in this case Windsor factor) is introduced to incorporate the quantum effects. 

\end{document}